\def\year{2022}\relax
\documentclass[letterpaper]{article} % DO NOT CHANGE THIS
\usepackage{aaai22}  % DO NOT CHANGE THIS
\usepackage{times}  % DO NOT CHANGE THIS
\usepackage{helvet}  % DO NOT CHANGE THIS
\usepackage{courier}  % DO NOT CHANGE THIS
\usepackage[hyphens]{url}  % DO NOT CHANGE THIS
\usepackage{graphicx} % DO NOT CHANGE THIS
\usepackage{natbib}  % DO NOT CHANGE THIS AND DO NOT ADD ANY OPTIONS TO IT
\usepackage{caption} % DO NOT CHANGE THIS AND DO NOT ADD ANY OPTIONS TO IT
\DeclareCaptionStyle{ruled}{labelfont=normalfont,labelsep=colon,strut=off} % DO NOT CHANGE THIS
\usepackage{xcolor}
\usepackage{array}
\usepackage{multirow}

\newcommand{\answerYes}[1]{\textcolor{blue}{#1}} 
\newcommand{\answerNo}[1]{\textcolor{teal}{#1}} 
\newcommand{\answerNA}[1]{\textcolor{gray}{#1}}

\newcommand\probjp{\texttt{pro-BJP}\ }
\newcommand\other{\texttt{Other}\ }

\newcommand\MyBox[2]{
  \fbox{\lower0.75cm
    \vbox to 1.7cm{\vfil
      \hbox to 1.7cm{\hfil\parbox{1.4cm}{#1\\#2}\hfil}
      \vfil}%
  }%
}
\usepackage{algorithm}
\usepackage{algorithmic}

\usepackage{subcaption}
\usepackage{newfloat}
\usepackage{listings}
\floatstyle{ruled}
\newfloat{listing}{tb}{lst}{}
\floatname{listing}{Listing}

\usepackage{threeparttable}

\title{Unraveling the Dynamics of Television Debates and Social Media Engagement: Insights from an Indian News Show}
\author{
    Kiran Garimella\textsuperscript{\rm 1}
    Abhilash Datta\textsuperscript{\rm 2} 
}
\affiliations{
    \textsuperscript{\rm 1}Rutgers University \\
    \textsuperscript{\rm 2}Indian Institute of Technology Kharagpur\\
    kiran.garimella@rutgers.edu,
    abhilashdatta8224@gmail.com 
}
\begin{document}

\maketitle

\begin{abstract}

% State the problem, your approach and solution, and the main contributions of the paper.

The relationship between television shows and social media has become increasingly intertwined in recent years. Social media platforms, particularly Twitter, have emerged as significant sources of public opinion and discourse on topics discussed in television shows. In India, news debates leverage the popularity of social media to promote hashtags and engage users in discussions and debates on a daily basis.

This paper focuses on the analysis of one of India's most prominent and widely-watched TV news debate shows: ``Arnab Goswami -- The Debate''. The study examines the content of the show by analyzing the hashtags used to promote it and the social media data corresponding to these hashtags. The goal is to understand the composition of the audience engaged in social media discussions related to the show.
The findings reveal that the show exhibits a strong bias towards the ruling Bharatiya Janata Party (BJP), with over 60\% of the debates featuring either pro-BJP or anti-opposition content. Social media support for the show primarily comes from BJP supporters. Notably, BJP leaders and influencers play a significant role in promoting the show on social media, leveraging their existing networks and resources to artificially trend specific hashtags.
Furthermore, the study uncovers a reciprocal flow of information between the TV show and social media. We find evidence that the show's choice of topics is linked to social media posts made by party workers, suggesting a dynamic interplay between traditional media and online platforms.

By exploring the complex interaction between television debates and social media support, this study contributes to a deeper understanding of the evolving relationship between these two domains in the digital age. The findings hold implications for media researchers and practitioners, offering insights into the ways in which social media can influence traditional media and vice versa.

\end{abstract}

\section{Introduction}

%\kiran{There are four main results until now: 
%1. There is a significant fraction of tweets which are being posted by BJP supporters.
%2. These tweets are specifically concentrated to specific topics and ignore other topics.
%3. The posting volume significantly increases during certain external events which show BJP in a negative light (not done yet, just my hypothesis)
%4. There is a clear connection between what the active BJP supporters tweet and what republic debates.
%}

%What is the problem?
%Why is it interesting and important?
%Why is it hard? (E.g., why do naive approaches fail?)
%Why hasn't it been solved before? (Or, what's wrong with previous proposed solutions? How does mine differ?)
%What are the key components of my approach and results? Also include any specific limitations.

Television and social media platforms have emerged as influential players in shaping public opinion and political discourse. The interdependence between these two spheres has garnered considerable attention in recent years, as they collectively contribute to the agenda-setting process and influence public perceptions~\cite{brojakowski2015television,valenzuela2017comparing}.

In this paper, we focus on the popular prime time debate show ``Arnab Goswami -- The Debate" on Republic TV, which is the most-watched English language news channel in India~\cite{broadcast_audience_research_council_data_2022}. Broadcasting from 9 to 11pm India time on weekdays, this show garners a significant viewership of over 5 million individuals across the country.
An intriguing aspect of this show is its active engagement with social media platforms. Each day, the show's producers promote a specific hashtag related to the debate topic, urging social media users to actively participate in discussing and deliberating upon the subject matter. This integration of television programming with social media interaction forms the basis of our investigation.

%we delve into the intricate relationship between the most watched Indian television news debate show, its social media support, and how their interplay benefits the political influence of the Bharatiya Janata Party (BJP). 
%Our study sheds light on the dynamics of this interplay, unraveling the mechanisms through which the show and social media support mutually shape and reinforce each other's narratives.

%We focus on the prime time debate show on Republic TV, the most watched English language Indian TV news channel, and link it with social media data about contents discussed on the show. The show, titled `Arnab Goswami - The Debate' airs every week day from 9--11pm India time and is watched by over 5 million viewers in India.
%The show makers promote a hashtag every day specific to the debate of the show and encourage social media users to debate the topic on social media.

By examining the content discussed on the show and analyzing the corresponding social media data, we aim to comprehensively understand the intertwined nature of television news debates, and social media support, and political influence, specifically in the context of the BJP in India. While previous research has explored the impact of media and social media on political communication~\cite{bimber2014digital,gerodimos2015obama}, there remains a dearth of studies that investigate the connection between television shows and a political party's online support base. Our study fills this gap by examining the content, audience composition, and influence of social media discussions surrounding the television news debate show closely aligned with the BJP.

We show that there is a clear bias in the topics chosen by the show and they favor the BJP significantly. 
Next, we show that the audience who engages with the show's content on social media is mostly made up of BJP supporters, with official BJP spokespeople also actively promoting the show.
These BJP supporters specifically amplify pro BJP and anti opposition narratives from the show that benefit their agenda.
Given the influence and infrastructure the BJP has on social media, this provides a significant boost to the show.
Finally, and most importantly, this relationship between the show and the party does not seem to be just one way. We find evidence that some of the hashtags being used on the show were previously used by BJP supporters, indicating that the show might be picking up topics to debate through content generated on social media.

This research is of paramount importance due to several reasons. Firstly, understanding the relationship between television news debates and social media support is crucial in comprehending the contemporary media landscape and its impact on political communication. The influence of partisan media on public opinion and agenda setting has far-reaching implications for democratic processes. Secondly, analyzing the BJP's utilization of social media support and its symbiotic relationship with the television show sheds light on the tactics employed by political parties to shape public discourse and maintain their political influence. Such political influence might be an age old practice typically done through back channels and lobbying, but our study provides some evidence of its existence.

Previous work in this space either look at the topics of discussion on the show~\cite{subhajit2021media} or on the influence of the news on social media~\cite{de2022indian}.
Such approaches that solely focus on either television news debates or social media discussions fail to capture the interdependence and feedback loop that exists between these two domains. By integrating content analysis of the television show, examination of social media support, and understanding the influence of political narratives, our study offers a comprehensive perspective on this intricate interplay.
Also, even though literature on political communication and media effects have talked about the impact of traditional and new media~\cite{guo2019social}, quantitative evidence of the specific feedback loop observed between television news debates and political party supporters may not have existed prior to our work.

Understanding these dynamics is essential for fostering informed public debate, promoting media literacy, and safeguarding the democratic ideals that underpin a healthy media ecosystem.

\section{Related Work}

\noindent\textbf{Use of hashtags on TV shows}. 
The use of hashtags on television shows has proven to be an effective means of connecting with viewers on social media platforms. By incorporating hashtags, TV shows can encourage audience engagement, facilitate discussions, and promote the show to a wider audience. This approach not only generates free promotion and marketing for the show but also fosters a sense of community among viewers, ensuring the longevity of the program~\cite{mukherjee2014social}.

%\textcolor{blue}{
Audience engagement has emerged as a vital aspect, especially with the rise of user-generated content~\cite{newman2009rise}. In leading news organizations, audiences are increasingly becoming active contributors to the way stories are researched and told. This lends credence to our findings regarding the strong role of party workers and supporters in shaping the discussion topics of the debate show we study in this paper. User-generated posts often lead news bulletins, pointing to a historic shift in control towards individual consumers and grassroots political actors.
%}

Scholars in the field of communication have long explored the relationship between television and social media, recognizing social media as a platform for re-articulating audience engagement. However, capturing and understanding audience engagement, especially with the proliferation of media channels, has posed challenges for researchers~\cite{moe2016rearticulating}.

The emergence of social media as a ``back channel" for television, as described in the book by~\cite{proulx2012social}, has opened up new opportunities for marketers to reach and engage audiences in innovative ways. The concept of ``Social TV" gained attention in the early 2010s, highlighting the potential for viewers to connect with others who share similar interests and engage in synchronous discussions during television viewing~\cite{wohn2011tweeting}. However, despite initial enthusiasm, many television shows have since abandoned the use of social media channels as a primary means of promoting their shows~\cite{hu2014social,guo2019social}.

Numerous television shows have embraced the use of hashtags to engage their audiences. For example, shows like \#TheEllenShow, \#Survivor (a reality game show on CBS), and \#TheTonightShow have encouraged fans to create and participate in discussions using specific hashtags. Popular television series such as \#GameofThrones and \#StrangerThings have also witnessed active hashtag-based conversations among dedicated viewers.
Several news shows, including CNN's Anderson Cooper 360 (\#AC360) and Fox News's Tucker Carlson Tonight (\#TuckerCarlsonTonight), utilize hashtags as a promotional tool. However, these news programs do not employ distinct hashtags for individual episodes or segments.
%While popular television shows such as `The Ellen Show', `Survivor', and `The Tonight Show' have successfully leveraged hashtags to engage their fan base, it is worth noting that many news shows in the United States, such as Anderson Cooper 360 (\#AC360) and Tucker Carlson Tonight (\#TuckerCarlsonTonight), do not employ specific hashtags for individual episodes or segments.

One notable application of hashtags in the television realm has been during US presidential debates, starting in 2014. \citet{robertson2019democratic} argue that social media hashtags have played a democratic role in spreading information about political debates, while~\citet{lin2013bigbirds} explore the social dynamics of emergent hashtags. \citet{khosla2019events} demonstrate that the use of debate hashtags on Twitter can even predict users' voting behavior.

%\textcolor{blue}{
Lastly, understanding the multifaceted relationship between media platforms and their publics adds another layer of complexity~\cite{coleman2010media}. Various publics are increasingly finding ways to subvert the gatekeeping functions of mainstream media to articulate their voice. This phenomenon deepens our understanding of the intricate dynamics we observed between the media, public discourse, and political engagement concerning the studied television debate show.

\noindent\textbf{Impact of social media on newsroom content}.
%\textcolor{blue}{
The existing literature on the interrelationship between traditional news media and social media platforms is extensive and multidisciplinary, covering communications, media studies, and political science. In recent years, both news organizations and social media platforms have found that their missions increasingly overlap, each moving into domains traditionally ascribed to the other~\cite{braun2011hosting}. This confluence resonates with our study, where we observe a reciprocal flow of information between the TV show and its social media audience. In similar lines, news organizations are grappling with the need to host user communities that may not abide by traditional journalistic norms, while social media platforms are dealing with content that resembles news and challenges their established user guidelines.
%}

%\textcolor{blue}{
Furthermore, the disruptive impact of social media on traditional news media's business models is another area of concern~\cite{newman2011mainstream}. This paper's findings extend the context of our analysis by highlighting the need to understand the commercial implications of the interplay between social media and television debates. News organizations are becoming increasingly cautious about the influence of social media on their economic sustainability, which complements our observation on how the popularity of television debates is bolstered by targeted social media activities, particularly from political supporters.
%}

The connection between social media and the choices editors make in newsrooms have also been studied previously in various other contexts.
\citet{sen2015clicks} study the causal relationship between social media popularity and how that influences editorial decisions in newsrooms. They show that editors prefer to schedule follow up content for stories which are popular on social media much more.
\citet{mukerjee2023metrics} also show similar trends for editorial choices for newsrooms to post content on Facebook. Editors typically prefer posting more content on popular topics which are being engaged highly because of the advertising dollars associated with engagement.
\citet{petre2020all} explores the dual nature of social media metrics in journalism. While the metrics serve as a form of managerial surveillance and can lead to increased pressure on journalists to improve their metrics, some journalists also find ways to leverage metrics to assert their professional value and autonomy, challenging the perception of metrics as solely disempowering. The book offers insights into data-driven journalism and offers a glimpse into how the metrics revolution could impact other professions in the future.

\noindent\textbf{Indian television debates and social media}.
Indian television debates have come under scrutiny recently due to their poor quality and biased nature~\cite{trial-by-media}. In the study by \citet{akbar2022devotees}, the authors demonstrate how television debates enabled the formation of social media communities to contribute to the politicization of a death by suicide. \citet{chakravartty2015mr,subhajit2021media} found that a majority of prime time debate shows in India exhibit a pro-government bias.
\citet{bhat2022expanding} conducted an analysis of television  shows in India and discovered that instead of facilitating meaningful deliberation on important civic issues, these shows often prioritize promoting religious majoritarianism, defending the policies of the Modi government, and advocating hyper-nationalism. The hosts of most TV debate shows employ a combative style and polarizing tone to stifle dissenting voices and impede free expression.

On social media, \citet{de2022indian} found that the BJP engages more with anchors and commentators from these debate shows. It is therefore expected that the support for these shows predominantly comes from an audience sympathetic to the BJP's agenda. Previous research~\cite{jakesch2021trend} also reveals the existence of an organized machinery of users on social media who actively tweet and amplify BJP hashtags by making use of a complex network of WhatsApp groups and pre-written templates to trend hashtags.

These findings shed light on the complexities and biases within Indian television debate shows and social media, highlighting the need for critical examination and fostering of more inclusive and balanced discussions.

%%%%% suggestions

%%%%% end suggestions

%The quality of debates has been terrible. \cite{trial-by-media} shows how biased.

%\cite{akbar2022devotees} show how social media communities enabled the politicization of a death by suicide.

%A majority of prime time debate shows have been pro government. \cite{chakravartty2015mr,subhajit2021media}

%Television debate shows in India have been a topic of discussion. \cite{bhat2022expanding} found that instead of enabling meaningful deliberation on a variety of important civic issues, television talk shows in India were fixated on promoting religious majoritarianism, defending the Modi government’s policies, and advocating hyper-nationalism. The hosts were found to stifle dissenting voices and forestall free expression using a combative style and polarizing tone. 

%\cite{de2022indian} shows that BJP engages more anchors/commentators.

%Since these shows mostly cover pro-bjp content, it is expected that their support also comes from an audience sympathetic to the cause.
%However, what we show is that there is an organized machinery of users who tweet and amplify the show. These users explicitly support the BJP or are .

%\cite{subhajit2021media} study debates on prime time and find that a majority of them are pro BJP.

%\kiran{self: find citations of the use of hashtags on tv}

\section{Datasets}

%\kiran{give some context of the importance of republic bharat and arnab.}

%Hashtags have become an integral part of social media and online discourse in India, and have also made their way into prime-time television shows. Popular news shows, including those on Republic TV, often display hashtags on-screen during discussions and debates, and encourage viewers to participate in the conversation online using the same hashtags. The use of hashtags allows viewers to engage with the show and share their opinions and perspectives on social media platforms like Twitter, Instagram, and Facebook. It also helps to create a sense of community and shared identity among viewers, and can amplify the reach and impact of the show's content.

%However, the use of hashtags on prime-time television shows has also been criticized for promoting polarization and divisive discourse, as it can encourage viewers to take sides and engage in online arguments and trolling. Additionally, the use of hashtags can sometimes lead to the spread of misinformation and fake news, as viewers may share unverified or biased information without fact-checking.

%Overall, the use of hashtags on prime-time television shows in India is a reflection of the growing importance of social media in shaping public discourse and political narratives, and underscores the need for responsible and ethical journalism that prioritizes accuracy, fairness, and impartiality. Henceforth, t

\subsection{Background}

%\section{Background}

Republic TV is an Indian English-language news channel founded by the journalist Arnab Goswami in May 2017. The channel has been the most-watched English news channel in India since its inception, according to data from the Broadcast Audience Research Council (BARC) with an average viewership of 40\%~\cite{broadcast_audience_research_council_data_2022}.
Republic TV is known for its sensationalization of news, and its controversial anchor Arnab Goswami who has been accused of being biased and of consistently taking a pro-hindu, pro-nationalist and pro-government tone~\cite{subhajit2021media}.

%In the week of January 10-16, 2023, Republic TV had a viewership share of 39.3\% in the English news genre, followed by Times Now with 23.4\% and CNN News18 with 18.3\%. Republic TV's viewership share in the super-primetime (9pm-11pm) band was 44.83\%, compared to 25.44\% for Times Now and 16.46\% for CNN News18.

%\subsection{Background}
%Republic TV is an Indian news channel that was founded in 2017 by journalist Arnab Goswami. It quickly gained popularity for its coverage of breaking news and controversial issues, and for its outspoken and opinionated style of journalism. Arnab Goswami, who is also the managing director and editor-in-chief of the channel, became a prominent figure in Indian media and politics, known for his aggressive interviewing style and nationalistic views. The channel has been involved in a number of high-profile controversies and legal battles, including allegations of biased reporting, sensationalism, and spreading fake news. Despite these criticisms, Republic TV remains one of the \textbf{most-watched news channels} in India.

\begin{figure}
    \centering
    \includegraphics[width=\linewidth]{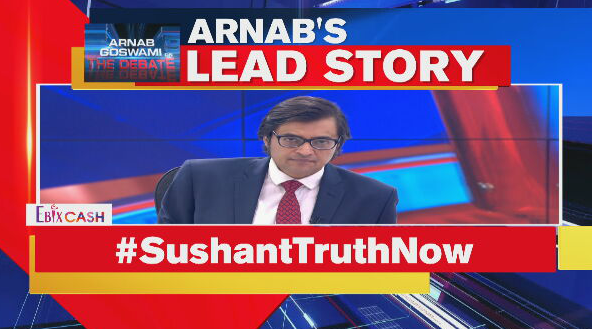}
    \caption{Example hashtag being promoted at the beginning of the show}
    \label{fig:example}
\end{figure}

Republic TV's flagship shows are the prime time debate shows hosted by Arnab Goswami at 9pm and 10pm every week day with over 5 million daily viewers. 
%The show airs every week day and depending on the news cycle, there is a prime time debate 2-3 times a night.
Every debate has a unique hashtag which is promoted by the host of the show. Since its inception, these debate shows have been using hashtags to promote discussion about the topic on social media. The anchors often remind viewers about using these hashtags to engage with them on social media.
Figure~\ref{fig:example} shows an example hashtag (\#SushantTruthNow) being promoted at the beginning of the show.

\subsection{Debate hashtags dataset}

We first start by collecting all debate shows that aired on Republic TV in 2022 by scraping their website.\footnote{\url{https://www.republicworld.com/the-debate}} To isolate the effects of Republic TV debates on Twitter users and to remove the influence of other news channels, common hashtags that were used by both Republic TV and other news channels were removed from the dataset. We also manually went over the hashtags and removed any generic hashtags which could be used outside the show. This removed 28 hashtags. We were left with 636 hashtags which we can be  sure that they were unique and used only by the show.

%This is achieved by collecting the user IDs of Twitter handles associated with other news channels, such as TimesNow, NDTV, and AajTak, and identifying the hashtags that are commonly used by both Republic TV and these news channels. These common hashtags are then removed from the dataset to obtain a clean and focused view of the impact of Republic TV debates on Twitter users.

Next, using the Twitter API V2, we searched for these hashtags and obtained all the users who tweeted about these hashtags.
We also apply a temporal filter and restrict our search to tweets that were created within a 3-day window before and after date the debate show aired using the hashtag. This additional step ensures that the tweets captured in the dataset are directly related to the debate show, and not due to other external factors. 

Typically, hashtags used in debates have a relatively short lifespan, lasting anywhere from a few hours to a few days. Most tweets start immediately after the 9pm show airs and have a lifespan of a few hours.
However, the longevity of a hashtag depends on several factors, including the popularity and relevance of the debate topic, the level of engagement from users, and the overall visibility of the hashtag across social media platforms. Some hashtags may continue to be used for several days or even weeks if they spark a broader public conversation or if they are picked up by influential social media users and amplified to a wider audience.

Next, we got the 3,200 most recent tweets for the users who have tweeted at least three of these hashtags. This cut off ensures that the users in our dataset were actively engaged in discussing the topics covered by the debates.\footnote{The limit of three hashtags was decided based on the practical limitations of the API. Due to the uncertainty of the availability of the API at the time of the data collection (March 2023), we chose a cutoff that enabled us to get the most amount of data in a reasonable amount of time.}
We also collected the profile information of these users.

Our final dataset contains 50,580 users with over 107 million tweets, of which 1.7 million contained hashtags used in the debates. 
90\% of the tweets were in either English or Hindi. Even though the show was only in English, we found roughly 40\% of the tweets by the users were in Hindi.\footnote{{
The language of the tweet was obtained directly from the Twitter API results~\cite{twitterEvaluatingLanguage}.}}

% \begin{figure}[t]
% \centering
% \includegraphics[width=]{users_vs_time.png}
% \caption{Distribution of users tweeting republic hashtags vs. time of year 2022}
% \label{fig:topic_model}
% \end{figure}

\subsection{Classifying BJP users}
One of our primary hypothesis is that there is a connection between the BJP members and Republic TV shows.
To study this connection, we need a classifier which can classify if a user is pro BJP or not.
To identify users who are supporters of the BJP, we developed a machine learning model. Given a Twitter user, the model identifies whether the user is a BJP supporter or not.\footnote{BJP supporters are broadly defined as supporters of causes that the party espouses, including Hindu religious accounts, and users who promote hyper-nationalism. The users who are not in the BJP supporter class could include a wide range of users who are either opposition party supporters or just news consumers.} In the rest of the paper, we refer to these two classes as \probjp and \other.
%We trained on a dataset of BJP supporters and opposition supporters/neutral users. that includes users who support BJP's agenda (broadly defined, hence including hindu nationalists too) and negative examples of BJP supporters in a supervised manner. Then we evaluate the model using a subsample of collected data (not trained upon) and check the classification reports. Finally, we manually test it on a small sample, to see how it is really performing.

%\subsection{Training Data}

To build a comprehensive dataset for analyzing political leanings on Twitter, we used two different types of sources. 
%\textcolor{blue}{
The dataset was annotated in a semi-supervised manner by leveraging the hashtags employed by users. We operated under the critical assumption that hashtags with established political leanings can serve as proxies for the user's own political bias. This approach allowed us to acquire a large-scale dataset efficiently. We manually validate this assumption in later in the section.
%}

\begin{enumerate}
    \item  Coordinated posting of party specific hashtags: 
For \probjp, we used the dataset from~\citet{jakesch2021trend}, who show that these users coordinate and post BJP talking points on Twitter to get hashtags trending. 
%\textcolor{blue}{
Jakesch et al. provide concrete evidence (see Section 7.1 in their paper) that these accounts were indeed affiliated with the BJP.%}
We obtained all tweets from these users.
For \other, we used hashtags by the Twitter account @trend4india, which actively promotes anti BJP hashtags every day.
%\textcolor{blue}{
Since there is no similar documented top-down operation run by the Congress party, we used the Twitter account @trend4india, which does not seem to be officially linked to the Congress but regularly promotes pro-Congress (and anti-BJP) hashtags and organizes trending raids similar to the BJP ones documented by ~\citet{jakesch2021trend}.
A user is designated as \probjp or \other based on their utilization of a minimum of three unique hashtags from the corresponding sets.%}

\item We used the dataset of politician accounts from ~\citet{panda2020nivaduck} to obtain BJP and all other politicians for \probjp and \other respectively. We used hashtags used by opposition politicians to obtain additional accounts for the \other category.

%\item We used hashtags which were non-political during October 2022 to obtain a list of user accounts for the \other category.

\end{enumerate}

\begin{table}[]
\caption{Number of users obtained from each dataset}
\label{tab:datasets}
\centering
\begin{tabular}{l|l}
\hline
\multicolumn{1}{l|}{}        & \multicolumn{1}{l}{\textbf{Num users}} \\ \hline
Coordinated posting (pro-BJP) & 4,220                                   \\
Trend4India (Other)           & 4,158                                   \\
Nivaduck Pro-BJP              & 3,690                                   \\
Nivaduck Other                & 3,780                 \\
\hline
\end{tabular}
\end{table}

%\textcolor{blue}{
Overall, this gave us 16k users (7.9k \probjp and 8k \other). We chose a stratified random sample of 12K users for training (6k of each class) and a stratified random sample of 4k users for validation (2k users of each class).
The final number of users we used in our classification (after the sampling) is shown in Table~\ref{tab:datasets}.%}
Next, we obtained the recent 3,200 tweets using the API for these users and pre-processed the data by removing stopwords, lower casing and lemmatization.

%\kiran{this is a bjp vs not bjp classifier. so it could encompass both anti bjp but also neutral participants. hindu supporters are classified as pro bjp}

We created term frequency-inverse document frequency (tf-idf) vectors of the tweet text.
We observed that mentions and hashtags particularly help in the classifier performance and hence created a separate vector of mentions, hashtags and urls for each user using  td-idf. Following the technique presented in~\cite{datta2022personality}, we pass these mentions/hashtags/urls vectors through a five layer fully connected deep neural network and train it. Once the model has been properly trained, we take the output of the second last hidden layer of the neural network and call it the hashtag/mention/URL embedding for each user. According to~\citet{datta2022personality}, this embedding captures the user's personality features well, much more so than text. 
%Overall, this approach allows us to analyze the political beliefs and affiliations of Twitter users in a more nuanced and sophisticated way. By leveraging the power of deep learning techniques, we can extract valuable insights from the mentions, hashtags, and URLs used in tweets, which can help to inform political strategy and decision-making.

For each user, we concatenated the embedding vector with the tf-idf vector from the tweets text and passed it to classical machine learning classifiers such as Random Forest Classifier, SVC, Logistic Regression, and XGBoost for classification.
We found that Random Forest performs the best, scoring an accuracy of 0.91 and an F1 score of 0.91 on the validation set. 
%\textcolor{blue}{
Table~\ref{tab:validation} shows the results on the best model.%}

\begin{table}[]
\caption{%\textcolor{blue}{
Classification report of the model on full validation set}%}
\centering
\label{tab:validation}
\begin{tabular}{lllll}
\hline
             & Precision & Recall & F1-score & Support \\
\hline
Other            & 0.93      & 0.89   & 0.91     & 1,600    \\
Pro-BJP            & 0.89      & 0.93   & 0.91     & 1,570    \\
accuracy     &           &        & 0.91     & 3,170    \\
%macro avg    & 0.91      & 0.91   & 0.91     & 3170    \\
%weighted avg & 0.91      & 0.91   & 0.91     & 3170   
\end{tabular}
\end{table}

After the validation, we also manually tested our model on 
 an unseen sample of 500 users. The testing data is selected randomly but is stratified to ensure that the class distribution is similar to the training and validation datasets. We obtain the predictions from the model on the testing data and then manually checked the user's profile to see the correctness of the classification. 
 %We mark the false positives and false negatives and investigate the reasons behind their wrong classification. This step is crucial as it helps us identify potential shortcomings of our model and ways to improve it. 
 The confusion matrix for the model from the manual testing is shown in Table \ref{tab:conf-matrix}. As we can see, the model performs reliably well on the task. We manually evaluated the small number of false positives and false negatives and found that the classifier confused journalists who mostly tweet for pro government stances (such as \texttt{@OpIndia\_com}) as \other. 
In most false positive cases, users have used the word `Hindu' in a non-political sense in some of their tweets.

\begin{table}[h]
\centering
\renewcommand\arraystretch{1.5}
\caption{%\textcolor{blue}{
Confusion Matrix on the test set}%}
\label{tab:conf-matrix}
\setlength\tabcolsep{5pt}
\begin{tabular}{l|l|c|c|c}
\multicolumn{2}{c}{}&\multicolumn{2}{c}{Prediction}&\\
\cline{3-4}
\multicolumn{2}{c|}{}&\probjp&\other&\multicolumn{1}{c}{Total}\\
\cline{2-4}
\multirow{2}{*}{Actual}& \probjp & $207$ & $31$ & $238$\\
\cline{2-4}
& \other & $9$ & $253$ & $262$\\
\cline{2-4}
\multicolumn{1}{c}{} & \multicolumn{1}{c}{Total} & \multicolumn{1}{c}{$216$} & \multicolumn{    1}{c}{$284$} & \multicolumn{1}{c}{$500$}\\
\end{tabular}
\end{table}

%\begin{eqnarray*}
%Precision &=& 0.976 \\
%Recall &=& 0.871 \\
%F1~Score &=& 0.921 \\
%Accuracy &=& 0.930 \\
%\end{eqnarray*}

%\kiran{also add some information on the manual testing and what the false positives/negatives were.}
%\ad{DONE}

%For the testing phase, the model was evaluated on an unseen sample of 300 data points, where all pro-Hindu Twitter accounts were considered as BJP supporters. As a result, some pro-Hindu accounts were classified as false negatives, as the model mistakenly labeled them as non-supporters of BJP. For example, the Twitter handle @KSKRISHNAN11 has a picture of the Hindu God Krishna as his profile picture, but the model failed to classify it as a BJP supporter. On the other hand, some Twitter users expressed their support for specific BJP members, such as Nupur Sharma, but the model failed to classify them as BJP supporters. 

%\subsection{Inference}
% Applying the classifier on our whole Republic Dataset. \textbf{50.5\%} supporters \textbf{plot}
Finally, we applied the classifier over the entire Debate Hashtag Dataset and found that 24,149 (47.74\%) of all users in the dataset are BJP supporters. 

In our analysis, we further divide the BJP supporters to identify \textit{active} BJP supporters as those who have tweeted more than 10 different hashtags of Republic debates.
These users are not only supporters of BJP but are also actively engaged in discussions and debates related to Republic TV and BJP. In our dataset, there were 2,065 such users.

\section{Analysis}

\subsection{Hashtag analysis}
\label{sec:hashtag_analysis}
We manually annotated all the 636 hashtags into different categories.
%\textcolor{blue}{
One of the authors conducted the annotation through an iterative process. Initially, we established coding categories for hashtags, drawing inspiration from work by~\citet{caravan_republic_tv_debates_bias}. We then started qualitative coding to identify emergent themes, refining these through successive iterations until they could be aggregated into final codes.
Most hashtags were unambiguously classifiable, but some necessitated group discussion among the authors for accurate categorization.
We instituted a policy that each hashtag must be assigned a single `primary' topic. In the few instances (less than 1\% of cases) where a hashtag could belong to multiple categories, we allocated it to one primary topic for the sake of analytical rigor.%}

Upon annotation, we found that a significant portion of the debates covered anti opposition topics (34\%),
with other popular categories were being covered by the show pro BJP (28\%), hyper-nationalism (9.5\%)\footnote{%\textcolor{blue}{
We specifically opted for the term `hyper-nationalism' over the term `nationalism' used by~\citet{caravan_republic_tv_debates_bias} because the traits expressed by the show --labelling anyone disagreeing with them as `anti-national' or or aggressively advocating for exclusionary practices that marginalize certain communities-- are better captured by the former term.}%}
, news (8.6\%), Russia (6.7\%) and Pakistan (3.5\%).
Only 2.1\% of the shows featured hashtags that were anti BJP.
This is in line with previous analysis which showed that the channel picks topics which are mostly pro government~\cite{caravan_republic_tv_debates_bias}.

%\subsection{Activity over time}
%who these BJP supporters and

\begin{figure*}[ht]
    \centering
    \includegraphics[width=\linewidth,trim=0 28 0 0,clip]{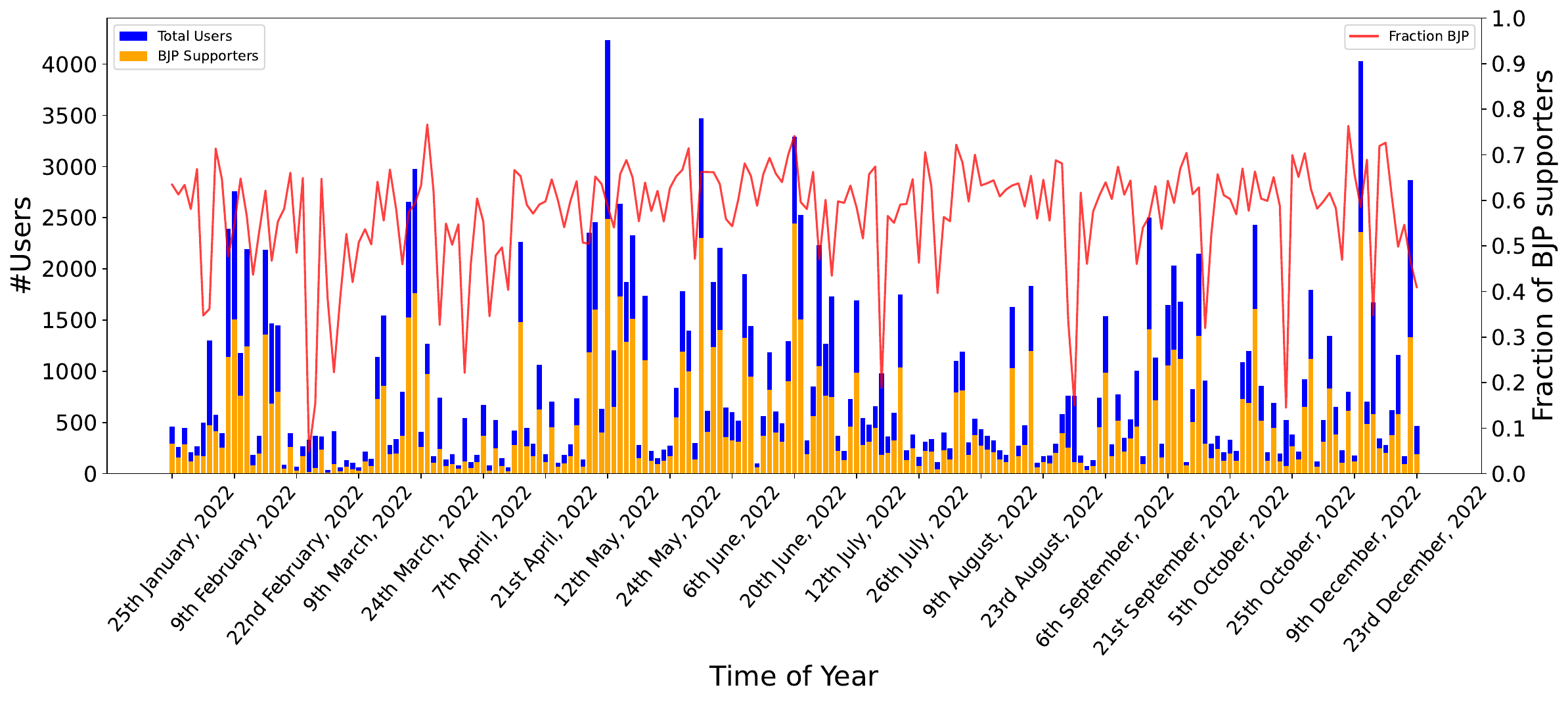}
    \caption{Number of users tweeting Republic TV debate hashtags over time. The blue bars show the total volume of users. The orange bars indicate BJP supporters. The red line (second y-axis) shows the fraction of the total users each day which was BJP supporters. We can observe that this fraction is consistently around 60\%.
    %\kiran{can we have a second y axis which shows the fraction of tweets which were by BJP supporters? Can you also point me to where the code is for these? in case I want to edit any minor stuff, like the font sizes etc? }
    %\ad{GIVEN}
    }
    \label{fig:users_vs_time}
    
\end{figure*}

Next, we analyze the activity of posting about these hashtags over time.
Figure~\ref{fig:users_vs_time} shows the number of users who tweet the hashtag over time.
We see that on average around 1,000-1,500 users post tweets about the debate every day.
There are a few periods of high activity which are analyzed  in Table~\ref{tab:peak-analysis}. 
%The dataset is representative of a diverse range of users who are active on Twitter. We train a classifier to identify which of these Twitter users are supporters of the BJP in Section~\ref{sec:}.
The red line (corresponding to the second y-axis) plots the ratio of the orange and blue bars, indicating the fraction of users on any day who are pro BJP. Interestingly, we see that this fraction is almost consistently around 60\%.

By analyzing the news around the peaks in Figure~\ref{fig:users_vs_time}, we want to understand whether there were any external events which drove an increase in engagement of the hashtags.
As we can see (in Table~\ref{tab:peak-analysis}), most of the peaks correspond to events which were popular politically and in many cases related to events where the BJP government was involved.

%Overall, our dataset provides a valuable resource for researchers interested in understanding the political landscape of Twitter users who are interested in the debates covered by Republic Bharat. The comprehensive data collected for each user in our dataset allows for detailed analysis, providing insights into the political views and opinions of users in our dataset.

%\kiran{Can we also look at how many of these 96k accounts are also present in the niva duck database? So we want to know how many politicians also tweet these hashtags.}

\begin{table}[h]
\begin{threeparttable}
\centering
\caption{Narratives for the peaks in Figure \ref{fig:users_vs_time}.}
\label{tab:peak-analysis}
\begin{tabular}{c|c}
%\multicolumn{2}{c}{\textbf{Peak Analysis}}\\
% \cline{3-4}
\hline
\textbf{Date Of Peak} & \textbf{Narrative} \\
\hline
8th February 2022 & Hijab ban\tnote{a}\\
18th March 2022 & Kashmir Files\tnote{b}\\
\multirow{2}{*}{14th April 2022} & Religious tensions due \\
 & to Ram Navami\tnote{c}\\
12th May 2022 & Gyanvapi mosque\tnote{d}\\
1st June 2022 & PFI Conspiracy\tnote{e}\\
\multirow{2}{*}{20th June 2022} & Opposition Government\\
& collapse in Maharashtra\tnote{f}\\
%20th June 2022 & opposition Government \\ 
% & collapse in Maharashtra \\
6th September 2022 & Bengal violence\tnote{g}\\
20th September 2022 & TRP report acquits Republic\tnote{h}\\
10th December 2022 & India/China clashes\tnote{i}\\
22nd December 2022 & Bharat jodo yatra\tnote{j}\\
\hline
\end{tabular}
\begin{tablenotes}
\item[a] \url{https://archive.is/DGg4V}
\item[b] \url{https://archive.is/MTYUr}
\item[c] \url{https://archive.is/31LXw}
\item[d] \url{https://archive.is/geY1u}
\item[e] \url{https://archive.is/ccOMl}
\item[f] \url{https://archive.is/PoKkI}
\item[g] \url{https://archive.is/2ECpn}
\item[h] \url{https://archive.is/d7Tw0}
\item[i] \url{https://archive.is/VSj4a}
\item[j] \url{https://archive.is/70pgt}
\end{tablenotes}
\end{threeparttable}
\end{table}

There seems to be a give and take relationship between the party and the channel, for instance, when Republic was acquited in a case, there was a huge support by BJP folks even though it was not related to BJP or politics.
Given how popular the show is, this finding raises questions about the influence of media on public opinion formation and the potential for media outlets to shape the narrative and priorities of political discourse.

\subsection{Users posting debate hashtags}

Next, we look at the users posting the debate hashtags by  looking at properties of twitter usage in the three sets of users (active BJP supporters, BJP supporters and other).

\begin{figure}[ht]
\centering
\includegraphics[width=\linewidth, trim=28 2 0 0, clip]{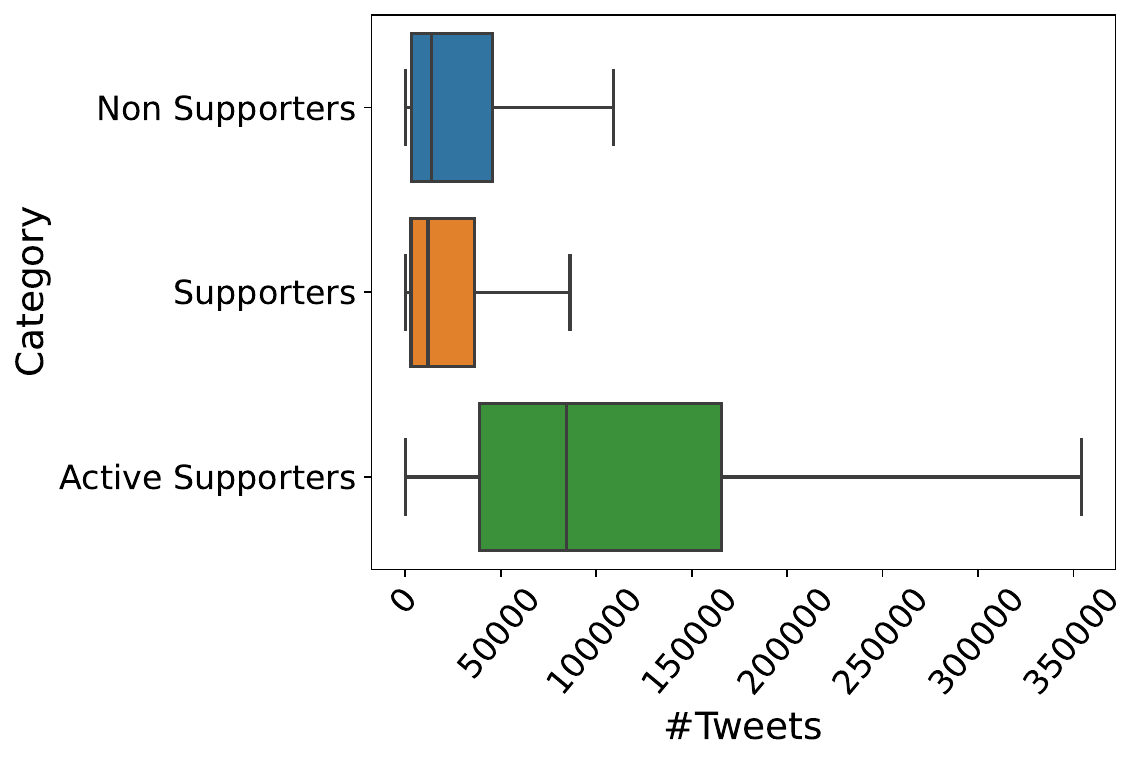}
\caption{Comparison of the number of tweets for the three sets of users.}
\label{fig:tweets}
\vspace{-\baselineskip}
\end{figure}

\begin{figure}[ht]
\centering
\includegraphics[width=\linewidth, trim=28 3 0 0, clip]{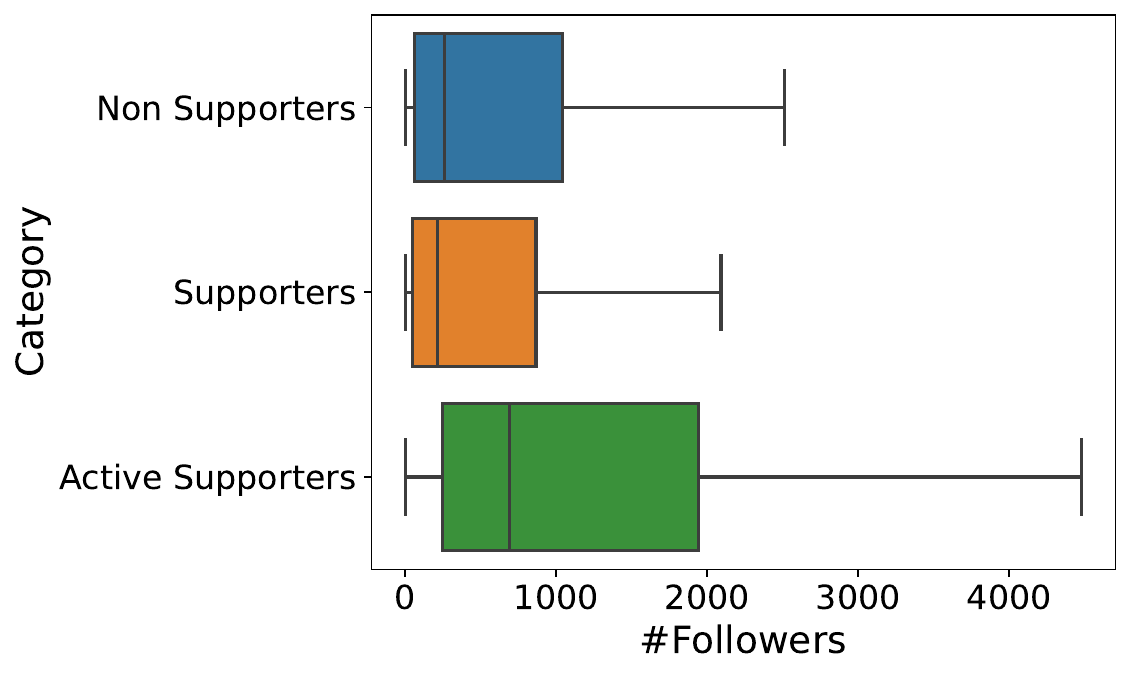}
\caption{Comparison of the number of followers for the three sets of users.}
\label{fig:followers}
\vspace{-\baselineskip}
\end{figure}

%Figure~\ref{fig:followers} shows the number of followers for the three groups. Active supporters have a lot more followers than the other groups.
Our analysis of tweet count and follower count reveals a significant difference between active supporters and the other two user sets (p$<$0.001). Active supporters have a considerably higher number of tweets and followers compared to the other groups, as shown in Figure~\ref{fig:tweets} and Figure~\ref{fig:followers}. Notably, the most followed and retweeted users within the active supporters group are predominantly BJP supporters or official spokespersons, such as @gauravbh, @Shehzad\_Ind, and @shaziailmi. This finding establishes a clear connection between the individuals involved in the discussions and the amplification of messages related to these debates.

Additionally, the most followed and retweeted users in the general supporters category also consist mostly of BJP supporters or affiliates. However, it is important to note that their messages have limited reach, as indicated by the median number of retweets, which is 20 for supporters compared to 7 for non-supporters. This suggests that the influence and dissemination of messages are more concentrated among the active supporters and their affiliated accounts.

%Analysis of the tweet count and follower count indicates that active supporters have significantly ($p$$<$0.001) more tweets and followers compared to the other two user sets (Figures~\ref{fig:tweets},~\ref{fig:followers}).
%The most followed and retweeted users in the active supporters group mostly include BJP supporters or official spokespersons such as @gauravbh, @Shehzad\_Ind and @shaziailmi. 
%This clearly reveals a connection between who engages in the discussion and amplifies messages related to these debates.
%The most followed and retweeted users in the supporters category are also mostly bjp supporters or bjp affiliates.
%
%The median number of retweets for supporters is 20 where as the non supporters is 7. So even their messages are limited in reach.

\subsection{What are they amplifying?}

\begin{figure}[ht]
\centering
\includegraphics[width=\linewidth]{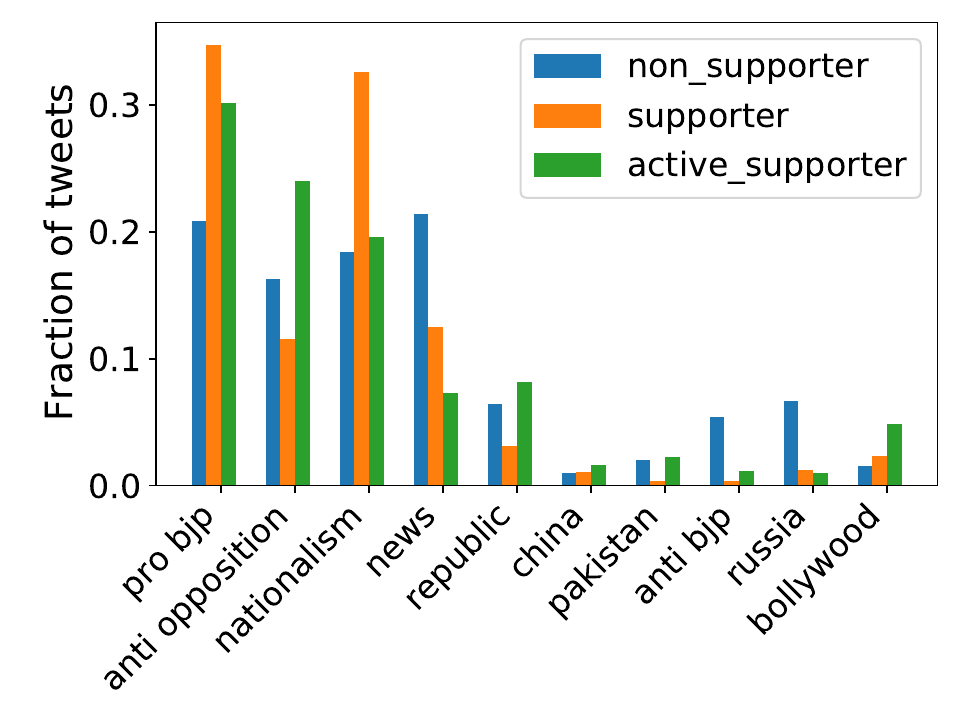}
\caption{Topics being amplified by the three sets of users..}
\label{fig:topics_fractions}
\end{figure}

The analysis reveals that these supporters selectively amplify content aligned with their own interests. Figure~\ref{fig:topics_fractions} provides insights into the preferred topics for each group, leading to several key findings:

\begin{itemize}

    \item Supporters predominantly amplify pro-BJP and hyper-nationalism related content. Their focus is largely centered around promoting the party and its ideologies.

    \item Active supporters, in addition to pro-BJP and {hyper-nationalism} content, heavily engage in discussions related to the opposition, often taking an anti-opposition stance. Interestingly, they also show a significant interest in Bollywood content, such as criticizing Bollywood celebrities.

    \item Non-supporters, on the other hand, tend to focus more on general news and a diverse range of topics, including issues related to Russia.
    
\end{itemize}

The findings indicate a notable presence of BJP supporters on Twitter who actively engage in discussions surrounding Republic TV debates. This highlights the distinct preferences and priorities observed among different user groups. Supporters demonstrate a strong inclination to promote their preferred political party and its ideologies, whereas non-supporters engage in a wider range of discussions encompassing diverse news topics and international affairs.

%They amplify the topics from Figure~\ref{fig:topic_model} much more than other, say, critical topics like crime or education.

\subsection{Coordinated Posting}

%\kiran{if finding the coordinated posting doesnt give us anything, lets try semantic embeddings. faiss should work extremely fast. we can just work with a small set of vectors just from active bjp users to start. e.g. https://github.com/facebookresearch/faiss/wiki/Faiss-building-blocks:-clustering,-PCA,-quantization}

Next, we looked at whether there was any coordinated posting posts promoting the debate hashtags.
\citet{jakesch2021trend} present how pro BJP hashtags are manipulated using a network of WhatsApp groups to coordinate posting and get hashtags trending on Twitter.\footnote{An example of coordinated posting: \url{https://archive.is/v9wfi} promoting the an initiative by the BJP.}
The manipulation works by using a set of template tweets which are copy pasted by a network of BJP supporters at a pre agreed time.
We explored whether any such coordinated posting exists in our dataset. We used fuzzy matching with Levenshtein distance to identify near similar tweets.
We were able to identify seven Republic debate hashtags which were manipulated this way in our dataset by BJP supporters.
The hashtags have specific themes like promoting development of Kashmir (\#TirangaAtLalChowk), supporting BJP government decisions (\#indiawithagniveer), supporting Hindu causes (\#hinduholocaust, \#kashmirfilestruth, \#justiceforharsha), supporting BJP victories (\#yogireturns), and amplifying anti opposition topics (\#banpfi). Though this only represents 1\% of all the hashtags we studied, this bolsters the evidence that there is a concerted connection between the hashtags the show uses and the hashtags official BJP accounts promote.
We also obtained data of what hashtags were trending on the day these shows aired and found that six of the seven hashtags were trending on the day they were manipulated.

%The full list of hashtags manipulated is prese
%Out of them only 6 were also used by Republic. `only' because even 6 is not low.

%\#kashmirfilestruth, \#harghartiranga, \#hinduholocaust, \#indiawithagniveer, \#yogireturns, \#banpfi, \#justiceforharsha have evidence of coordinated posting. 

%We identified hundreds of hashtags which are coordinated and manipulated this way. 
%Using near similar tweets using fuzzy matching (with levenstein distance).

%\kiran{TODO for self: add more analysis of trending hashtags}

%This list may not be complete because XXX, we found that overall 43 of these hashtags were trending.

\subsection{What else do they tweet?}
\label{sec:topics}

The tweets containing republic hashtags only make up roughly 1.5\% of the tweets by the users in our dataset. What else do the users in our dataset tweet? This would help us learn more about these users and the topics they post. To identify these, we applied topic models on the users tweets.
%
%\textcolor{blue}{
To identify topics, we first selected a tweets from each user category: active supporter, supporter, and non-supporter. We then employed Latent Dirichlet Allocation (LDA) for topic modeling on these sets.%}
We use the LDA implementation provided by Gensim~\cite{vrehuuvrek2011gensim} to obtain topics from our tweets. The LDA model generates a probability distribution over the topics for each tweet, indicating the likelihood of the tweet belonging to a particular topic. It also returns the most significant words belonging to a particular topic.
We experimented with various topic numbers ranging from 10 to 150 and picked the one (N=50) which gave us the most coherent topics. 
%\textcolor{blue}{
We experimented with smaller numbers of topics, but found that they were insufficient to capture the richness and diversity of themes in the dataset. Topics were either too broad or overly fragmented, making it challenging to gain meaningful insights. On the other hand, when we tested with a higher number of topics, say 150, the topics became too granular and often exhibited significant overlap, making it harder to interpret and use the results effectively. After careful examination, we found that 50 topics struck a balance, providing a coherent and interpretable representation of the underlying structures within the data without being overly detailed or redundant.%}
We run separate topic models for the three different sets of users (active supporters, supporters and non supporters).

After generating 50 topics for each dataset, we manually label them with high-level categories. We chose 10 different high-level categories to label the topics. These are politics, religion, crime, business, health, hyper-nationalism, international, terrorism, bollywood and other.  We ensure that the labeling is consistent across all the different user sets.\footnote{This was needed to be able to compare topics across the different user sets. However, we had to make a compromise and choose high level topics which had to be aggregated at a level which matches all three user sets.} To label the topics, we follow a two-step process. First, we analyze the most important words within each topic and identify the underlying theme. Then, we assign a high-level category to each topic based on the underlying theme. For example, if a topic has frequent words as `BJP', `Modi', and `Manifesto', we assign it to the `politics' category, and if a topic has frequent words as `Hindu', `Shiva', and `Muslim', we assign it to the `religion' category. If some topic, doesn't match any of our high-level categories, we assign it `other' category.
We found that the topics we came up with in Section~\ref{sec:hashtag_analysis} were too narrow for this analysis. This is expected since the users might tweet many other topics outside of the topics covered by the debates.

\begin{figure}
\centering
\includegraphics[width=\linewidth, trim=0 27 0 5, clip]{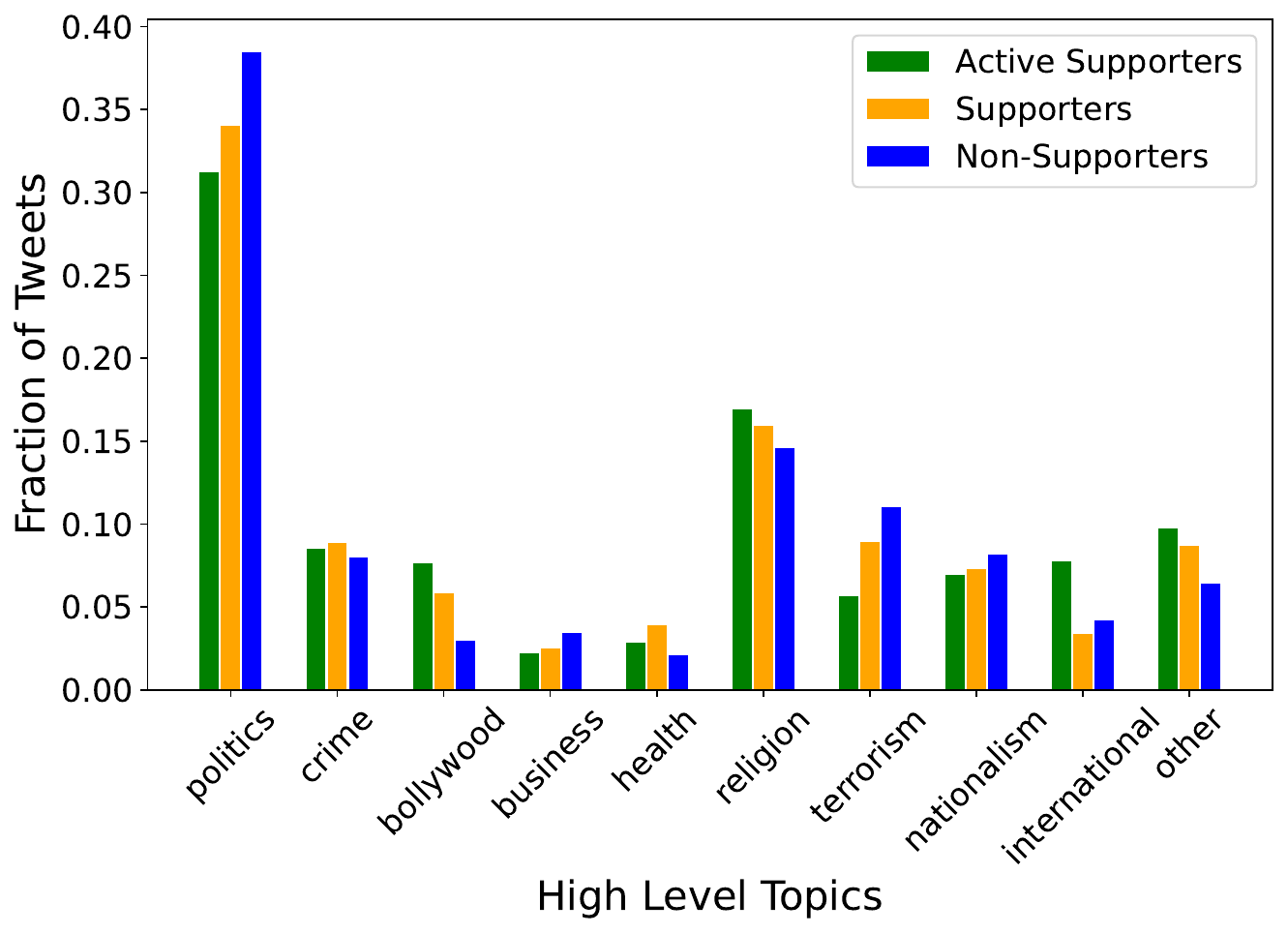}
\caption{Comparison of tweet topics for the three user sets.}
\label{fig:topic_model}
\end{figure}

Once we labeled the topics, we calculate the counts of tweets in each high-level category in our dataset. 
%We define the support of a category as the number of tweets that belong to that category. 
We combine all counts of topics belonging to a specific high-level category to find its total count. 
This provides insights into the most prevalent topic categories among the different user sets. A comparison of the tweet counts of different categories between the three user types is shown in Figure~\ref{fig:topic_model}.

On comparing the tweet distributions of the three user types, we see that non supporters tweet much more about Politics and issues around terrorism.
Supporters and active supporters tweet more on religion and Bollywood. This rhetoric of supporters of the BJP consistently targeting particular Bollywood celebrities and movies has been documented previously~\cite{kommiya2022voting,akbar2022devotees}. 
Overall, our analysis closely aligns with previous literature. It is expected that active BJP supporters would be highly engaged in political discourse and religious matters, while non-BJP supporters would be more concerned with social and political issues. 

\subsection{Impact of social media on debate hashtag choice}

\begin{figure}
    \centering
    \includegraphics[width=\linewidth]{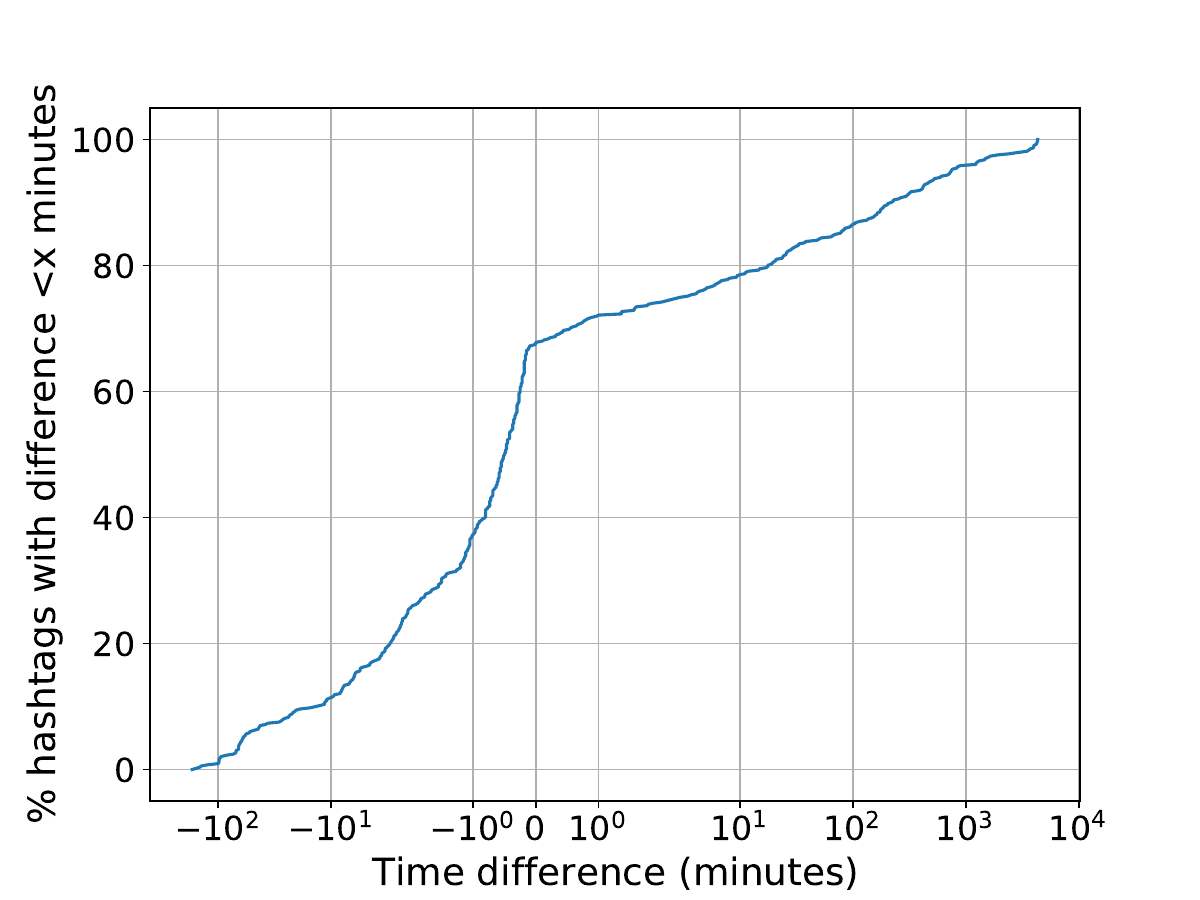}
    \caption{CDF of the time difference (in minutes) between when a hashtag was first shared on social media and when @republic or @republic\_bharat tweeted the hashtag.}
    \label{fig:cdf_difference}
\end{figure}

Finally, we study whether the hashtags chosen by the show are in any way connected to user's activity on social media. To study this, we check whether the hashtags used by the show were used prior by anyone on social media.
Since most of the hashtags used by the show are peculiar and unique, the chances of them being used earlier, even by BJP supporters should be rare.
Concretely, we computed the following: for each hashtag  used by Republic, we get the time when it was first posted by any of Republic's Twitter accounts (denoted by $t^h_{TV}$) and the first usage of the hashtag by a BJP supporter (denoted by $t^h_{SM}$).
We compute difference $t^h_{TV} - t^h_{SM}$. If the difference is negative, this means that Republic posted the hashtag first. If the difference is positive, a BJP supporter posted the hashtag first before it appeared on Republic TV.
Figure~\ref{fig:cdf_difference} shows a CDF of this difference over all the hashtags. 
We can see that around 35\% of the hashtags were used on social media prior to being used by Republic TV. Around 15\% of the hashtags were used at least 2 hours before Republic accounts used them.
%Qualitative analysis of these hashtags shows that some of them are popular hashtags generally e.g. \#LizTrussOut, \#WhereIsZelensky and \#NuclearThreat.

%If this diff is less than zero this means a certain hashtag was used by a supporter before it appeared on TV. 

%\kiran{please check and implement this. should be simple.}

Qualitative analysis of these hashtags reveals interesting observations. Take the case of the hashtag \#AbuseNo92, a hashtag keeping count of the number of abuses by the Congress party on Prime Minister Narendra Modi.\footnote{\url{https://archive.is/y1XBm}} The hashtag is very specific and there is a little chance that someone else had tweeted it earlier to it being chosen. The first tweet for this hashtag was done at 4pm India time by a user @SMedia4\footnote{\url{https://archive.is/OcUqG} The user claims on their bio `My TL reaches Modi ji ... '} and then there was a show using the hashtag at 10pm.
A few more examples of this pattern include: (i) \#HinduphobiaIsReal tweeted 3 days prior to the show using the hashtag (by a user who mostly tweets anti Congress topics), 
(ii) \#KashmirFilesPolitics was tweeted 2 days earlier by a user who claims to be `Proud to be an Indian, BJP supporter',
%https://twitter.com/PiyushShimpi1/status/1503324030685913094 (calling himself proud Indian and BJP supporter)
(iii) \#BanPFINow being used by a user 2 days earlier by a user who says in his bio `Luv my India and Narendra Modi Sir'.\footnote{All three users have less than a thousand followers. To protect the privacy of these users, we do not include their screen name.}

\begin{figure}
    \centering
    \includegraphics[width=\linewidth, trim=0 30 0 10, clip]{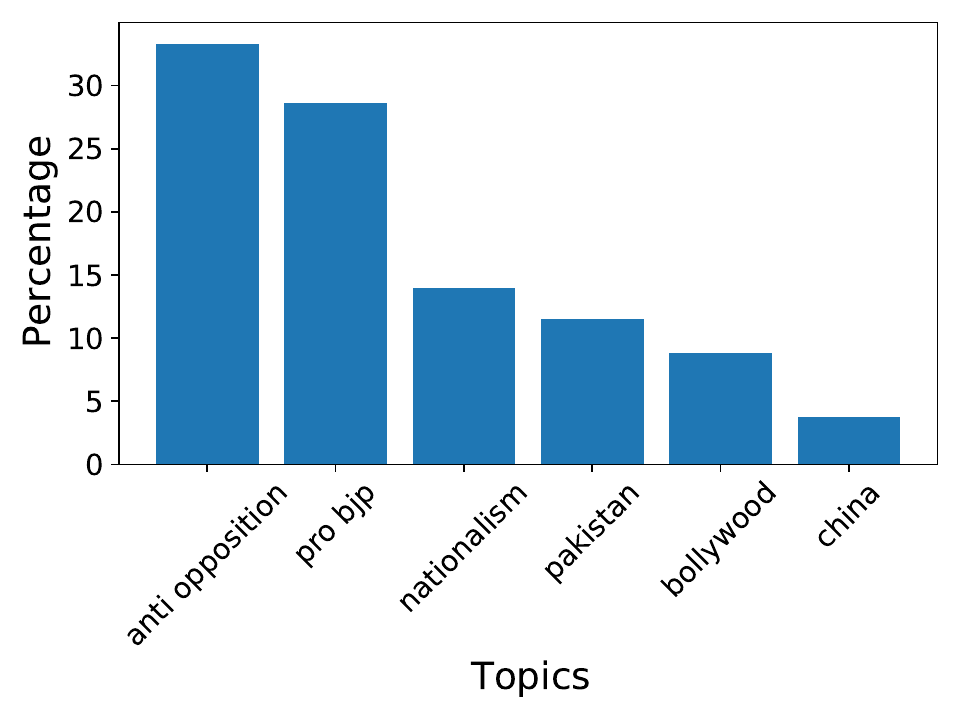}
    \caption{Topics posted earlier}
    \label{fig:topics_posted_earlier}
\end{figure}

We aggregated the topics which were posted earlier by social media users and Figure~\ref{fig:topics_posted_earlier} shows  that most of them were anti opposition or pro BJP hashtags.
Note that it is important to note that we do not claim that there is a causal connection here since identifying a causal connection is not possible given our data.
However, given that close to a 100 hashtags in the past year seem to have been used by BJP supporters prior to being used on the show does provide a signal that the show might be taking inspiration from social media users.

To find further evidence of this, we used topics labeled in Section~\ref{sec:topics} and created a time series of topic fractions for each of the three user sets per week. For each week, we compute the fraction of tweets belonging to a certain topic. We also labeled all the hashtags used by Republic into the same topics and created time series of the shows by Republic per week.
Next, we applied Granger Causality tests~\cite{granger1969investigating} on these time series to check if any of the topic time series can help predict the Republic TV show time series.
Granger Causality test is a statistical test that is used to determine if a given time series and it’s lags are helpful in explaining the value of another series.
If a certain time series $t_1$ `Granger causes' another time series $t_2$, it statistically means that the values in $t_2$ lag the values in $t_1$ by a certain amount and that $t_1$ can be used to predict values in $t_2$.
We set a lag of 4 weeks and run Granger Causality tests for all the 10 topics for the three user sets. Using a $p<$ 0.01, we find that only three of the time series have statistically significant values. These are topic time series for Bollywood and %\textcolor{blue}{
hyper-nationalism for active supporters and %\textcolor{blue}{
hyper-nationalism for the supporters group. %\textcolor{blue}{
This analysis provides more evidence that certain types of topics used by BJP supporters are correlated with hashtag choices made by Republic TV.%}

%Figure~\ref{fig:cdf_difference} shows the difference. 
%Most of these hashtags, particularly around the topics listed in Figure~\ref{fig:topics_posted_earlier} were tweeted first by either BJP supporters or active BJP supporters. This is clear evidence that Republic chooses hashtags (at least in part). Though this does not establish causal evidence, there is clear evidence.

%https://twitter.com/VisionaryRoopak/status/1572134791398686722 3 days earlier
%\#RespectHindusToo again a bjp talking point.
%Over 150 such hashtags.

%\kiran{is there no heuristic to remove these generic hashtags? they pollute our data.}

%\kiran{Next line of analysis could be on granger causality. We create the time series of topics (e.g. num tweets per week) posted by republic and by bjp supporters (active or not) and then apply tests like this \url{https://www.machinelearningplus.com/time-series/granger-causality-test-in-python/}}

\section{Discussion}

The study reveals that social media platforms, particularly Twitter, have emerged as influential channels for public opinion and discourse on issues discussed in television shows. The popularity of these platforms has been effectively harnessed by news debates in India, as they actively promote hashtags and encourage users to engage in discussions and debates on social media. 
%This highlights the increasing intertwining of television shows and social media support. 
Though this nexus between social media and television is over a decade old~\cite{proulx2012social}, there is not much study in the context described in this paper --- using social media to study the bias of a news show.

The analysis of hashtags used by the news debate show indicates a significant bias towards the ruling Bharatiya Janata Party (BJP). Over 60\% of the debates featured content that either favored the BJP or criticized the opposition. 
%The significant bias towards the ruling BJP observed on one of the most popular news debate shows highlights the channels ability to steer public discussions in favor of . 
By dominating the content and narrative of the show, the BJP effectively controls the information flow and influences the issues that are highlighted and discussed. This enables the party to set the agenda and shape public opinion in alignment with its own political interests.
While the initial impression of the finding may seem ``trivial", its significance becomes apparent when considering our subsequent two findings.

%This observation raises concerns about the impartiality and objectivity of the show, as it suggests a potential alignment with the ruling party's narrative. The prevalence of such bias in a widely-watched TV show underscores the significance of critically examining the influence of media on public opinion and the potential impact on democratic discourse.

Our next significant finding is the role of social media support in amplifying the show's reach and promoting specific hashtags. It is evident that the show garners substantial support from BJP supporters on social media platforms. BJP leaders and influencers actively participate in promoting the show, leveraging their existing networks and resources to artificially amplify and trend specific hashtags~\cite{jakesch2021trend}.

Finally, completing the cycle, we identify the reciprocal flow of information between social media and the topics picked for debate on the show.
%is an intriguing aspect uncovered by this study. 
Our evidence suggests that the show's choice of topics is influenced by social media posts made by party workers.
While causality cannot be established in this particular setup, our results suggest a low probability of this occurrence happening by chance. This is notably influenced by the active participation of BJP supporters in promoting the channel's content and the adoption of hashtags by influential users within their ranks.

%This phenomenon highlights the interplay between traditional media and online platforms, where political actors utilize social media as a tool to shape public discourse and extend their influence beyond the television screen. Such strategic efforts to influence the online narrative raise questions about the integrity of public opinion and the potential for astroturfing or the manipulation of not just social media trends, but opinions on television for political gains.

These findings highlight the dynamic interplay between traditional media and online platforms, where political actors utilize social media as a tool to shape public discourse and extend their influence beyond the television screen and in turn social media discussions can influence the agenda-setting process of television debates. 
This nexus means that the BJP sets a narrative on social media which is picked up and `debated' (amplified) by the TV show giving it legitimacy and prime time coverage, while the same BJP supporters amplify the debate narrative on social media promoting it to be an issue of national importance.
%So BJP sets a narrative which is taken up by Arnab, etc and then the same BJP supporters amplify that narrative as being national debate material..
It underscores the need to consider the complex inter-dependencies and feedback loops between these domains when analyzing the media landscape in the digital age.

Understanding the importance of our findings in the agenda setting by the BJP is crucial for several reasons. 
Firstly, our study documents the extent to which the relationship between the show and the party exist. It sheds light on the strategies employed by political parties to control the narrative and influence public opinion. 
%The BJP's dominance in agenda setting through the news debate show and social media support reveals the party's efforts to shape the public discourse in its favor. 
It is important to note that such agenda setting on media might already be happening through back channels (such as personal connections or lobbying) and are not usually visible. Our study provides some quantitative evidence of this connection.

Secondly, it underscores the need for media literacy and critical engagement among the public. Awareness of the biases present in media outlets and the influence of political actors on agenda setting is essential for individuals to make informed judgments and engage in meaningful debates.

Lastly, our findings call for further research and scrutiny into the role of political parties in agenda setting and the implications for democratic processes. The ability of a political party to control the agenda and manipulate public discourse raises concerns about the diversity of voices and perspectives being represented in the media landscape. It prompts discussions on media ethics, journalistic independence, and the role of media regulators in maintaining a balanced and inclusive media environment.

\subsection*{Limitations and Broader Impact}

While the study provides valuable insights into the relationship between a TV news debate show and social media support, there are several limitations that should be acknowledged:

Generalizability: The study focuses on one specific TV news debate show in India, which may limit the generalizability of the findings to other shows or contexts. The dynamics observed in this particular case may not be representative of the entire media ecosystem in India or other countries.
Even for India, English news viewers make up a small fraction of the viewership of news. Given the prevalence of media bias in multiple countries, it would be interesting to see if similar patterns exist for other countries or languages.

Sample Selection Bias: The analysis of social media data relies on a specific selection of hashtags related to the TV show. This selection may introduce a bias by excluding other relevant discussions or perspectives that do not use those specific hashtags. On a similar note, the topic of the show and its leaning/bias is decided based on the hashtag and not the content of the show which may also introduce bias in our analysis.

Choices in user classification: Our analysis quantifying BJP support for the show hashtags depends on an automated classifier. Even though the classifier performs well in terms of accuracy, there are false positives, which might bias the results. Our results must be interpreted with a margin of error that could include these false positives.
Our classifier identifies whether a Twitter user is a BJP supporter or not. Non supporters are not a homogenous group and contain a wide range of users from diverse ideologies, including neutral users. A finer way of classifying users should be thought of in the future.
%\textcolor{blue}{
Another issue which could impact our results is the misclassification. We thoroughly went through these examples of false positives and false negatives to understand where our classifier gets it wrong and tried to include features that capture such behavior. We found some cases where the classifier systematically gets it wrong. 
Our manual analysis indicates that the misclassifications are mostly towards classifying pro BJP as Other (the other way was quite rare).
In most of these cases, the accounts belonged to journalists or activists who support the BJP, but also tweet about other news related stuff. With these cases in mind, we tried to understand the downstream impact of our findings. Since our paper mostly focuses on pro-BJP support to the TV show, if anything, our results are slightly under counting the support.%}

%\textcolor{blue}{
In grappling with the ethical, political, and interpretive challenges inherent to our methodology, we acknowledge the complex landscape of decisions that come into play. Data labeling is not a neutral act; the very act of categorizing users as `supporters' or `non-supporters,' for instance, could be interpreted as an oversimplification of a complex and nuanced spectrum of public opinions. Additionally, the allocation of topics to tweets and hashtags is intimately tied to the politics of representation, often reflecting latent power structures and sociopolitical orientations within the media and broader society. To mitigate these issues, we employed a rigorous process that involved consulting multiple sources of information, including academic literature and collective authorial discussions, especially in cases where the categorizations were ambiguous. We also established guidelines to ensure that each hashtag or topic would be allocated to a single `primary' category, thereby maintaining analytical consistency and reducing interpretive bias.%}

%\textcolor{blue}{
On another note, our positionality in this research is anchored in a profound commitment to shed light on the anti-democratic tendencies manifest in the media landscape in India. As scholars, we are concerned with how specific media outlets may exert disproportionate influence in shaping public opinion, often in ways that align with the interests of ruling political parties, thus undermining the essential democratic dialogue. This informs not only the subject of our research but also the interpretive lens we adopt in analyzing our dataset. It is crucial to acknowledge this positionality to fully appreciate how our research questions were formulated, how the data were evaluated, and how our findings contribute to larger, ongoing debates about media ethics and the role of democracy in a digitized world.%}

%which clubs over 50\% of the users into the rest bucket. This bucket is not homogenous and includes supporters of other political parties, but also a majority of users who are just news consumers and need not specifically express their beliefs about any party.

Causality and Directionality: While the study suggests a reciprocal flow of information between the TV show and social media, it does not establish causality or the direction of influence. The observational setup we have in this paper does not allow us to make causal claims. Further research would be needed to determine whether the potential direction of influence we point out is actually a causal link.
%That is, even though there is quantitative evidence that certain hashtags were picked and used by Republic TV debates after the hashtags were tweeted by BJP supporters, this could have happened completely by chance. Our results just show the prevalence of this happening as non trivial.
%Though some of the results indicate towards a causal link, we would like to stress that our results do not provide enough evidence for a causal connection .

%Temporal Considerations: The study may be subject to temporal limitations, as media dynamics and social media trends can change over time. The findings may reflect the specific period of data collection and may not capture potential shifts or developments in the media ecosystem.
%For instance, Tucker Carlson tonight was the most watched show on American cable news. However, in May 2023 the show was cancelled and Carlson moved to social media to host his show. The viewership dynamics with this change.

%\noindent\textbf{Ethics statement}

Twitter focus: The study focuses on Twitter as a social network for convenience reasons. However, since Twitter is used only by elites in India, not focusing on other platforms like WhatsApp or Facebook might bias our findings to a certain type of user.
If Twitter is an elite social network with only a small fraction of the Indian electorate on it, why is a party like the BJP focused on creating infrastructure on Twitter? While the overall user base on Twitter may be smaller, it often includes influential individuals, journalists, opinion leaders, and celebrities who have a wide reach and can shape public opinion. Engaging with this influential subset of users can have a ripple effect on broader public discourse.

%-- even if this is not being done elsewhere, these strategies are copied and could be replicated.

%-- focus on twitter. why does it matter? Is this just a twitter phenomenon? or is it also happening on Facebook and WhatsApp? since Twitter is mostly an elite social network in India, why does it matter? (or maybe that is exactly why it matters).

%-- classification choices

%What ethics man?

%\section{Conclusion}

%It would be interesting to compare these with popular TV shows in the US where daily hashtags about the show (e.g. \#AC360 or \#TuckerCarlsonTonight) promote the show.

%There should be proper analysis done and a potential for causal link between social media and debate hashtag choice.

%\clearpage
% Use \bibliography{yourbibfile} instead or the References section will not appear in your paper
\bibliography{aaai22}

\end{document}